# Investigating the Online Recruitment and Selection Journey of Novice Software Engineers: Anti-patterns and Recommendations

**Miguel Setúbal · Tayana Conte · Marcos Kalinowski · Allysson Allex Araújo**

**Abstract** [Context] The growing software development market has increased the demand for qualified professionals in Software Engineering (SE). To this end, companies must enhance their Recruitment and Selection (R&S) processes to maintain high quality teams, including opening opportunities for beginners, such as trainees and interns. However, given the various judgments and sociotechnical factors involved, this complex process of R&S poses a challenge for recent graduates seeking to enter the market. [Objective] This paper aims to identify a set of anti-patterns and recommendations for early career SE professionals concerning R&S processes. [Method] Under an exploratory and qualitative methodological approach, we conducted six online Focus Groups with 18 recruiters with experience in R&S in the software industry. [Results] After completing our qualitative analysis, we identified 12 antipatterns and 31 actionable recommendations regarding the hiring process focused on entry level SE professionals. The identified anti-patterns encompass behavioral and technical dimensions innate to R&S processes. [Conclusion] These findings provide a rich opportunity for reflection in the SE industry and offer valuable guidance for early-career candidates and organizations. From an academic perspective, this work also raises awareness of the intersection of Human Resources and SE, an area with considerable potential to be expanded in the context of cooperative and human aspects of SE.

Miguel Setúbal
Federal University of Ceará (UFC) - Crateús *Campus*
E-mail: migueljonas@alu.ufc.br

Tayana Conte
Computing Institute - Federal University of Amazonas (UFAM)
E-mail: tayana@icomp.ufam.edu.br

Marcos Kalinowski
Department of Informatics - Pontifical Catholic University of Rio de Janeiro (PUC-Rio)
E-mail: kalinowski@inf.puc-rio.br

Allysson Allex Araújo
Center for Science and Technology - Federal University of Cariri (UFCA)
E-mail: allysson.araujo@ufca.edu.br

# 1 Introduction

In the context of a dynamic software market and the need for delivering high-quality products, organizations maintain the expectation of recruiting professionals adept in comprehending the fundamental principles and practices of Software Engineering (SE) [44]. The process of forming a workforce capable of fulfilling these roles necessitates the strategic Recruitment and Selection (R&S) of professionals, extending to individuals in the early stages of their careers, such as trainees or interns [79]. Within the contemporary employment landscape of the software industry, there is a demand for recruiting recently graduated or soon-to-graduate engineers for different reasons, including financial resources, talent formation, and team building [26, 62, 69, 70].

Moreover, the progression of digital transformation has prompted substantial modifications in professional R&S processes [23]. Notably, there is a trend towards the adoption of online recruitment and the emergence of recruiters who specialize in identifying candidates within the technology and software development sectors [34]. Consequently, organizations are increasingly leveraging the online environment to facilitate the dialectical process of mutual attraction described by Porter *et al.* [60]. In this process, candidates utilize the internet to explore job opportunities and present their professional experiences, while recruitment professionals exploit available resources and tools (such as LinkedIn, GitHub, Gupy, etc.) to assess potential candidates.

The R&S processes encompass phases, assessments, and sociotechnical factors [3]. In particular, these processes pose a substantial challenge for recent graduates aspiring to embark on a career in SE, given that each phase necessitates distinct experiences. Examples include participation in technical interviews, assessment of soft skills, cultivation of a digital presence, portfolio development, market awareness, and more. Consequently, we hypothesize the existence of unfavorable behaviors exhibited by early-career software engineers, which are recurrent from recruiters' perspective and impede a successful online hiring process. This issue assumes critical importance for both organizations, which contend with resource wastage, and candidates, who encounter frustration owing to the challenge of systematically preparing and enhancing themselves for entry into the professional environment [36, 39].

Considering the aforementioned rationale, we advocate for the implementation of an ongoing analysis tailored specifically for novice SE professionals. This initiative aims to address the nuanced intricacies of online R&S processes, providing recent graduates with the tools to proactively navigate potential social and technical drawbacks in the R&S context. The necessity for such particular discussion arises from the current scarcity of evidence-based guidance for entry-level developers. While extensive research in Human Resources and Management has explored R&S methodologies [3,36,42,53], a research scarcity still remains around empirical studies within the SE domain. This gap is particularly significant considering the unique characteristics of the SE industry, highlighting the need for targeted investigations to enhance the understanding and optimization of R&S practices in this specific domain.

Given the identified research gap, it becomes pertinent to initially discern the *anti-patterns* manifested during the online R&S processes involving early career professionals in SE. As *anti-pattern*, we draw inspiration from its conventional definition in SE: typical responses to recurring problems that are usually ineffective and can be highly counterproductive [18]. Under this analytical motivation, the primary objective of this study is to empirically investigate the following research question: *"What anti-patterns are committed by early career professionals undergoing the online R&S processes for positions in the field of Software Engineering?"*. The relevance of this research question lies in unraveling the pitfalls inherent in the current online R&S practices, offering a discussion that can inform the refinement of these processes for the benefit of both candidates and organizations.

To address our research question, we employed an exploratory and qualitative research design underpinned by data gathered from six Focus Groups (FGs). Each FG comprised three professionals, resulting in 18 participants from nine distinct companies, with substantial expertise in the R&S of software engineers. Concerning data analysis, we employed open and axial coding techniques [76], following the methodological guidelines proposed by Miles and Huberman [51]. The anti-patterns identified from the extracted data were also synthesized hierarchically through a comprehensive and interactive mind map[1]. Beyond elucidating these anti-patterns, we also offered actionable recommendations to alleviate them.

This research delivers contributions to industry and academia. Firstly, the mapping of 12 anti-patterns, coupled with 31 actionable recommendations, presents a valuable resource for SE research and practice. Early career candidates, in particular, stand to benefit from the recommendations provided, gaining a nuanced understanding of anti-patterns to avoid in the online R&S process. From the company's perspective, our analysis may contribute to optimizing talent acquisition strategies. Secondly, our study advances the broader academic discourse by delving into the intersection of HR and SE, an underexplored domain in empirical SE research. By bridging together these disciplines, we contribute to developing a more holistic understanding of the complexities involved in the R&S of novice software engineers. This interdisciplinary approach offers theoretical and practical implications that can inform future research endeavors in the intersection of HR and SE. Furthermore, our study acknowledges the behavioral and technical dimensions inherent in the R&S processes. This recognition aligns with the growing emphasis on considering social and human aspects in SE research, adding depth to understanding these elements intertwined with technical processes and their impact on the formation of highly qualified software development teams.

---

[1] Available at: https://gesid.github.io/papers/swe-novice-rs

This paper is structured as follows: Section 2 outlines theoretical concepts. Section 3 analyzes related work. Section 4 details the research design. Section 5 presents the results and analysis. Section 6 promotes an overall discussion. Section 7 addresses the limitations, and Section 8 offers concluding remarks.

**2 Background**

2.1 Recruitment and Selection Processes

According to the seminal work by Porter *et al*. [60], individuals and organizations exist in a perpetual and unceasing dialectical process. These entities partake in an ongoing and interactive process of mutual attraction. Similar to how individuals attract and select organizations by accumulating information and forming opinions, organizations aim to attract individuals and amass information about them to determine their suitability for employment.

Within a company, the oversight of personnel management typically falls under the purview of the Human Resources (HR) department, tasked with executing various processes related to HR management, prominently including recruitment and employee selection as integral functions [39, 53, 63]. As depicted in Figure 1, the **Recruitment and Selection** (R&S) processes can be summarized into four primary phases [3, 39, 77]. Specifically, **recruitment** encompasses the *attraction* and *screening* phases, while **selection** involves *assessment* and *hiring*. However, it is noteworthy that R&S are integral components of a unified process: the integration of new employees into an organization [66]. The rectangles in Figure 1 shaded in gray signify the R&S processes stages involving participant interaction. In contrast, the orange-shaded rectangles correspond to internal steps within each phase.

When a company needs to hire a new collaborator, the first step is to start the **recruitment** process. Recruitment refers to the process of finding individuals who are interested in the job vacancy offered by the company [78]. This process is responsible for communicating the need for personnel hiring within the company. The recruitment involves defining specific job requirements and screening candidates who meet the established profile [3]. Indeed, the recruitment process can vary in execution depending on the specific needs of an organization. However, its objective is clear: to attract a substantial pool of candidates with the necessary characteristics and competencies for a given position, forming the foundation for subsequent selection process [39].

As part of the recruitment process, the *attraction* phase involves several steps. Firstly, it is important to validate the need for a new employee by conducting a proper analysis of the organization's requirements and determining the staffing needs [66]. Once all staffing needs have been identified, it is relevant to define the candidate's profile based on the job objectives they should fulfill [78]. This issue requires a thorough understanding of the organization's requirements, providing directions into the desired candidate profile to fill the mapped vacancy [3].

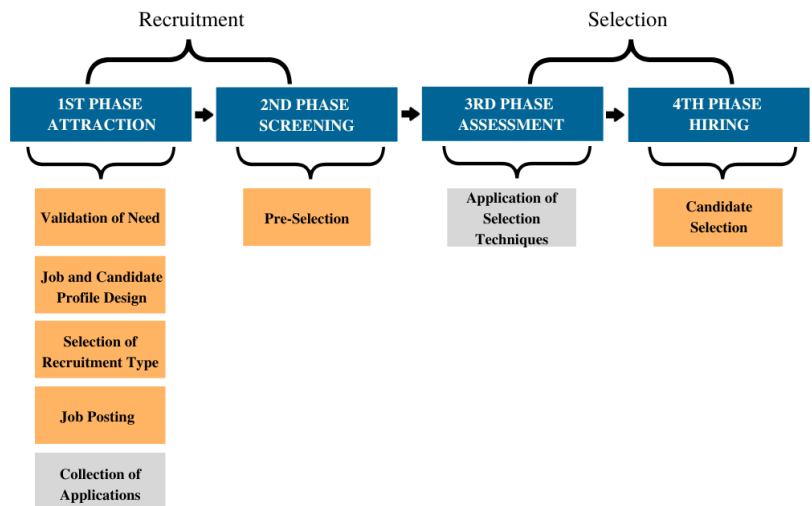

Fig. 1: Overview of the Recruitment and Selection (R&S) processes.

The organization must determine whether candidates should be sourced internally, by restructuring and reorganizing the existing workforce, or externally [39]. Internal recruitment predominantly involves seeking professionals already within the company, resulting in participants being transferred or promoted to new positions. In contrast, external recruitment stems from the need to explore candidates outside the organization [66].

Upon determining the recruitment method, the subsequent step involves promoting the availability of the requisite position. Accordingly, the vacancy must be advertised across diverse communication channels to reach the broadest possible audience, encompassing internal and external channels [39]. Job postings should be meticulously crafted to be ideal and enticing. This process is reciprocal, as the selection of new employees depends on both the company's choice of candidates and the candidates' choice of the company [64].

Furthermore, the application process typically initiates with the submission of a Curriculum Vitae (CV), a document detailing the candidate's educational and professional qualifications, potentially accompanied by a cover letter. The cover letter articulates the candidate's motivation for the position. It is relevant to emphasize that HR professionals should supervise the entire process to monitor its progression [3]. In contemporary times, technology has become an integral aspect of daily work, reflecting individuals' personal and professional image [27]. Digital platforms and social networks have emerged as means to enhance and streamline processes, facilitating swift comprehension and response to the needs of organizations and markets [59]. Consequently, organizations are increasingly leveraging social networks and digital platforms as external sources for recruitment, presenting an avenue to reduce costs, enhance processes, and expand opportunities for both companies and candidates [33].

After the job advertising, the next step is collecting applications for the positions to be filled. The more candidates, the better for the organization, as a larger pool of options provides the opportunity for a competitive selection where the most qualified candidate would be chosen [78]. When sending a CV or applying for a position on online platforms (*e.g.*, LinkedIn), candidates commit to either accept the company's offer or, at the very least, explore it, as well as the intended role [3, 77].

Following the collection of applications, which marks the conclusion of the attraction phase, the *screening* phase begins. The primary objective of screening is to overview and discern which candidates will progress to the subsequent process, namely, the selection [39]. At this point, the emphasis is on analyzing the material gathered from the vacancy posting. This step is also where the examination of CV takes place, aiming to identify the minimum requirements for the profile envisioned by the company and pinpoint candidates eligible for the next steps. Alongside this screening phase, candidates meeting the criteria established in the preceding stages are selected and subsequently introduced into the selection process [3].

The **selection** process follows the recruitment process. With the completion of the recruitment, a cohort of individuals possessing profiles most aligned with the available position is defined [64]. Consequently, a more rigorous process is necessary to determine which candidate will be selected for the job position. The selection process can be defined as the choice of the most suitable candidate for the organization from all the recruits, through various instruments for the analysis, assessment, and comparison of data [66]. In essence, the selection process ensures that organizations can choose individuals suited to a specific occupation. This process does not always entail selecting candidates with high skills or aptitudes; instead, it involves choosing candidates who closely align with the organization's expectations and exhibit potential based on predefined job requirements [19]. The selection process can be categorized into two phases: *assessment* and *hiring*.

In the *assessment* phase, pre-selected individuals undergo various assessments based on existing selection techniques that best align with and identify the competencies and potential of the candidates following the predefined job profile. Typically, the assessment phase incorporates the following techniques: CV (or resumes) analysis, tests, and interviews [66]. In the initial screening phase, participants' CV/resumes are examined to ascertain their conformity with the fundamental prerequisites of the position. Subsequently, during the assessment phase, a more comprehensive analysis is undertaken to extract pertinent information from candidates' resumes. This detailed analysis serves as a foundation for other types of assessments, predicated on individual attributes encompassing knowledge, skills, and attitudes [66].

After completing CV/resume analysis, candidates may undergo additional evaluation methodologies, including social and technical assessments. These tests aim to gauge the knowledge and skills acquired during the learning process [78]. Typically, these tests measure the level of professional and technical proficiency requisite for a given position. A diverse array of tests exists, encompassing oral, written, or practical formats, with each company selecting those aptest for assessing candidates *vis-à-vis* the specific position [66]. In summary, the essence of employment testing lies in utilizing a test that accurately predicts job performance. Hence, tests constitute a critical issue, with their outcomes determining which candidates proceed to the interviews.

The interview is one of the most influential techniques for determining the candidate's qualification [64]. Interviews provide an opportunity to delve into the candidate's prior experience and offer evidence of their interests and social attributes to the position [78]. The interview process, with its exploratory approach and non-standardized format, presents a unique opportunity to evaluate intangible qualities of job candidates, such as their motivation, resilience, and potential for making meaningful contributions to the organization. This approach makes it a valuable platform for assessing individuals beyond what can be measured through standardized employment forms [42].

Finally, after all the evaluation steps have been completed, decisions must be made based on all the collected information to determine the final candidates, *i.e.*, those who will be hired [78]. The last phase, *hiring*, should be carried out by the members of the R&S processes with the support of the organization's leaders for the available position. Both parties need to work together since a more informed decision can be made only with the opinion of the recruitment professionals who have the necessary knowledge about the candidates and the leaders who have a deeper understanding of the role's particularities.

2.2 Professional Practice in Software Engineering

The domain of professional practice in Software Engineering (SE) encapsulates the skills and attitudes requisite for software engineers to proficiently, responsibly, and ethically apply software engineering concepts in their projects [30,71]. Within this framework, SE professionals confront technical challenges inherent in software development, ensuring the delivery of reliable and functional software products. The accomplishment of this endeavor necessitates the acquisition of suitable skills, training, and hands-on experience, thus equipping individuals with the proficiency to effectively apply software engineering principles and methodologies in their projects [15]. As discussed by Meade [50], software engineers embody professionals who systematically apply disciplined and quantifiable approaches to software development, operation, and maintenance, encompassing a spectrum of responsibilities and functions.

The initiation of a professional career closely aligns with the development of new competencies to navigate the multifaceted challenges of the SE job mar-ket [32]. This trajectory starts with acquiring foundational knowledge during undergraduate courses [26]. The expansive field of SE comprises diverse subdomains, each corresponding to distinct stages of the software development process and, consequently, engendering varied specialties. These subdomains encompass, for example, requirements analysis, design, implementation, testing, and maintenance. Specialized roles have emerged within SE, including software architects, quality assurance engineers, DevOps engineers, and software project managers. However, it is relevant to emphasize that the titles and roles assumed by SE professionals within an organization exhibit substantial variability contingent upon its type and size.

Therefore, our objective here is to clarify the diversity of trajectories a novice SE professional can explore within the job market. This exploration encompasses various roles and specializations and delves into the distinct organizational contexts that shape professional pathways in SE. Recognizing and navigating this diversity becomes paramount for early-career professionals seeking to position themselves within the dynamic landscape of software engineering strategically.

## 3 Related Work

The section offers an overview of related work that delves into various aspects of professional R&S processes within Information Technology (IT) and Software Engineering (SE). These works contribute valuable discussions, ranging from the practices and challenges faced in recruiting and selecting IT professionals to the stages of the hiring process at specific IT companies. The papers were organized into three major groups: **Overall R&S efforts in IT/SE, Attraction/Screening phase, and Assessment phase**.

Concerning overall **R&S efforts in IT/SE**, Tyler [79] explored steps to build a successful software engineering team, offering insights into recruitment, hiring, and management. The provided a comprehensive guide for forming and managing teams in competitive environments by challenging recruitment myths and discussing industry-tested approaches. More recently, Ramkumar and Rajini [63] researched an effective R&S system for the Indian IT Software Industry. Statistical analyses revealed the need for tailored R&S models to address industry challenges and improve effectiveness. In addition, Swamy, Beloor, and Nanjundeswaraswamy [77] examined staffing, selection processes, recruitment sources, and employee satisfaction in the IT sector. Their study, based on responses from Bangalore IT companies, highlighted associations between R&S methods and organizational productivity, aiding decisions to enhance efficiency and satisfaction.

Regarding **Attraction/Screening phase**, Ehlers [22] evaluated the recruitment efforts of IT companies in the context of SE, emphasizing sociability in job postings. To this end, they analyzed data from job posting websites and feedback platforms containing organizational reviews quantified companies' efforts in recruiting qualified software engineers and emphasized the social benefits of advertised positions. In addition to standard benefits such as healthcare, vacations, and retirement savings, many companies highlight their sociability in job postings, explicitly listing offers of a supportive team, social events, or game rooms. Still addressing IT job opportunities, Montandon *et al.* [54] analyzed the distribution of hard and soft skills required by IT job opportunities. They focused on high-level hard skills across different developer roles. They approached a heatmap to visualize the distribution of skills and conducted manual annotation of soft skills from job descriptions. Their findings are valuable for understanding the skill requirements in the industry. Fritzsch *et al.* [24] investigated the phenomenon of overemphasizing trending technologies in job offerings and resumes within the software industry. Through an empirical survey involving software professionals in hiring and technical roles, the research uncovered the influence of Résumé-Driven Development (RDD) on the recruitment process. The findings highlighted the impact of trending technologies on job attractiveness for applicants and potential future employers. This work contributed to understanding the implications of RDD and raises awareness of its potential systemic effects on the software industry. In a follow-up study, Fritzsch *et al*. [25] analyzed the tendency to overemphasize technology trends in resumes and job applications, leading to potential negative consequences for both applicants and companies. Practical recommendations were provided to address this issue, underscoring the importance of focusing on core competencies, clearly communicating job requirements, and avoiding false expectations. They clarified the need for a balanced approach in showcasing skills and selecting technologies based on project-specific requirements rather than just following trends.

Lastly, covering the **Assessment phase**, Odeh and Tariq [58] discussed that the most effective way to measure the quality of potential employees is by testing candidates through coding assessments. However, this activity requires experts to evaluate the code and conclude the recruitment process when dealing with numerous potential candidates. Indeed, manual assessment becomes overly complex when domain expertise is unavailable, or there are numerous candidates. Consequently, the authors aimed to propose a system for assessing code quality using source code metrics and quality factors. The system was evaluated using code submissions from five developers, with the analysis revealing which code exhibited higher quality and which developer understood the importance of software quality. Still related to this perspective of coding assessments, Wyrich *et al.* [82] focused on constructing a theory to predict the influence of individual characteristics on performance in solving coding challenges. Their research involved an exploratory quantitative study where participants completed coding challenges and questionnaires. The findings revealed correlations between affective states, personality traits, academic performance, programming experience, and coding challenge performance.

Moreover, Behroozi *et al*. [9] focused on measuring the cognitive load of candidates during technical interviews conducted on whiteboards. Using a head-mounted eye-tracker and computer vision algorithms, the researchers found that candidates experienced higher cognitive load and shorter attention lengths when solving problems on the whiteboard than on paper. This research spotlighted the potential impact of interview settings on candidate performance and suggested the need for interventions to create a more inclusive technical interview process. Since 2018, Mahnaz Behroozi and Chris Parnin have been conducting different studies covering this perspective of technical interviews, including the stress influence [10,14], use of eye tracking systems [7], developers perceptions [11], asynchronous interviews [12], and bad practices [8, 13].

The previous studies discussed in this section served as references for exploring the scope of professional R&S processes in IT and SE. Although these works are relatively recent, there are still many challenges that need to be addressed. In essence, these works examine the particularities associated with online R&S of IT and SE professionals with a considerable part focusing on technical assessment. However, we identified no studies explicitly focusing on identifying anti-patterns or recommendations suggested by recruiters for early- career professionals in the SE industry. This opportunity is significant for organizations struggling with recurring issues in the R&S processes, as well as entry-level candidates who want to enhance their competitiveness in the professional environment.

## 4 Methodological Procedures

The R&S processes can be broken down into various phases, varying according to each organization. Each phase demands different qualifications from the candidates, who may engage in unfavorable actions that impact the recruiter's perception. Therefore, it is imperative for software engineers, especially those in the early stages of their careers with little (or no) market experience, to continuously seek best practices to guide them with the phases involving R&S. Hence, the complex dynamics of R&S, intertwined with the nuanced behaviors of novice engineers and their influence on recruiter perceptions, constitute a loosely investigated phenomenon. In this context, the adoption of an exploratory research, as suggested by Stebbins [74], emerges as a reasonable choice for a comprehensive investigation.

Given the need to understand sociotechnical elements of the phenomenon and its essence, this research adopts a qualitative research trajectory [65]. Regarding the procedures, an empirical perspective was established using online Focus Groups (FGs) to support investigations, counterpoints, and aggregations to assess facts and phenomena as they occur in reality [73]. In particular, we opted for FGs with professionals formally involved in software engineers' R&S process, with a view to promoting a planned discussion and obtaining a plural perception [5, 40].

Following the well-established guidelines for data condensation, data display, and drawing/verifying conclusions proposed by Miles and Huberman [51], the methodological procedures of this research were organized into six main steps, as illustrated in Figure 2.

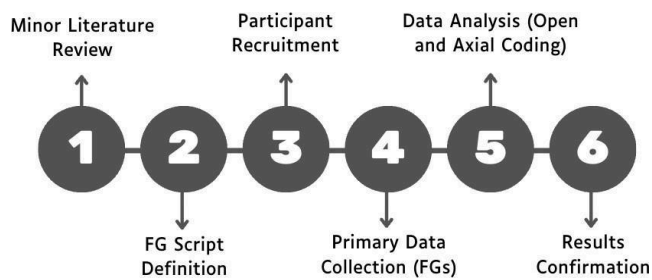

Fig. 2: Overview of the Methodological Procedures.

In the **first step**, a minor literature review [35] was conducted to understand the fundamental concepts and definitions of theoretical elements that link the fields of SE and HR (specifically, online R&S). For this review, a search was carried out by two authors on Google and Google Scholar between March and September 2021, using the following keywords: "software engineering recruitment and selection", "software engineering recruitment", "software engineering selection", and "human resources in software engineering". Our review primarily covered articles published in scientific conferences/journals, reports, and books written in Portuguese or English.

In the **second step**, we defined the interview script (available in our supporting repository [67]) for primary data collection through FGs (to be detailed in the fourth step). This script consisted of a semi-structured questionnaire organized into three main sections. Firstly, a brief overview of the research context to be presented to the participants. The second section contained open-ended questions regarding the anti-patterns and recommendations associated with each phase of the R&S process. We also included questions to cover general R&S aspects. Finally, the third section was designed to conclude the discussion, clarify doubts, and collect final feedback from the participants.

The **third step** involved inviting participants to join the FGs in the form of online meetings [73]. Two criteria were used for participant selection: i) they should be professionals in technology-based companies and ii) they should have more than one year of experience in recruiting and selecting software engineers, whether as recruiters, tech recruiters, managers, or tech leaders. The search for potential participants was conducted in three moments. Initially, potential participants were identified through snowball sampling using LinkedIn and the authors' professional network of contacts. Subsequently, individual invitations were sent via private messages or email. These invitations included a brief project description, including its risks and benefits, data collection steps, estimated duration, and the Informed Consent Form (ICF) for participating in the research. Along with the ICF, a questionnaire covering personal and professional/academic information was included to characterize the participants. Finally, the dates for online video conferences were scheduled in line with the participants' availability. It is worth noting that this study was conducted following the ethical principles and regulations established by the Ethics and Research Committee (CEP) of the Federal University of Ceará (UFC). The research protocol was submitted to CEP for evaluation and received approval under protocol number 5.726.955 on 10/27/2022.

Table 1 summarizes the characteristics of the selected participants, highlighting their professional profiles and academic backgrounds, among other information (the names and companies of the participants have been omitted to preserve confidentiality). On average, participants have 5.7 and 3.3 years of experience in traditional and online R&S, respectively. The most experienced participant has 20 years in traditional and 11 years in the online context. While two participants have no experience in traditional R&S, they have two years of experience in online R&S. The participants' academic backgrounds are also diverse, covering technology, psychology, and engineering. The FGs included individuals from the software development and HR fields, which induced a more plural and complementary discussion.

Table 1: Invited participants characterization

| FG | Part. | Higher Degree Course | Current Position/Role | Traditional R&S Exp. | Online R&S Exp. |
|---|---|---|---|---|---|
| FG1 | P01 | Psychology | People & Culture Analyst Tech Recruiter | 6 years | 3 years |
| | P02 | Systems Analysis and Development | Software Engineer | 2 years | 1 year |
| | P03 | Engineering | Head of Engineering | 1 year | 4 years |
| FG2 | P04 | Psychology | Recruiter | 11 years | 4 years |
| | P05 | Psychology | People and Culture Specialist | 20 years | 11 years |
| | P06 | Industrial Automation | Product Manager/Tech Lead | 2 years | 2 years |
| FG3 | P07 | People Management | People and Culture Analyst | 3 years | 3 years |
| | P08 | Psychology | People and Culture Analyst | 4 years | 4 years |
| | P09 | Computer Science | Head of Data | 2 years | 2 years |
| FG4 | P10 | Technology in Human Resources Management | Recruitment & Selection Analyst | No experience | 2 years |
| | P11 | Psychology | Tech Recruiter | 4 years | 3 years |
| | P12 | Administration and Organizational Psychology | Chief Executive Officer | 13 years | 6 years |
| FG5 | P13 | Technology in Human Resources Management | People and Management Analyst | 8 years | 2 years |
| | P14 | Psychology | People & Mgmt. Coordinator | 7 years | 3 years |
| | P15 | Computer Science | Executive Technology Manager | 7 years | 4 years |
| FG6 | P16 | Business Administration | People Manager | 10 years | 2 years |
| | P17 | Human Resources | People and Culture Analyst | No experience | 2 years |
| | P18 | Computer Science | Software Engineer/Scrum Master | 3 years | 3 years |

Building on the interview script, the **fourth step** focused on conducting the online FGs with the confirmed participants. Using the Google Meet tool and with the participants' consent, six FGs were conducted and recorded, each consisting of three individuals. The first occurred on 12/23/2022, lasting approximately 1 hour and 18 minutes. The second FG was held on 03/06/2023, lasting about 1 hour and 3 minutes. The third FG took place on 04/27/2023, lasting around 1 hour. The fourth FG was conducted on 05/17/2023, lasting approximately 1 hour and 10 minutes. The fifth FG was held on 05/12/2023, lasting about 1 hour and 3 minutes. The sixth and final FG occurred on 06/06/2023, with a duration of approximately 41 minutes. In total, the 18 participants contributed to 6 hours and 15 minutes of discussion, totaling 128 pages of transcription (in Times New Roman, font size 12).

While asking questions, the moderator (first author of this paper) aimed to provoke participants to explore and delve into the factors underpinning their responses, such as reasons, feelings, opinions, and beliefs [41]. It is important to note that the moderator should not have a position of power or influence but should encourage all types of comments, whether positive or negative [40]. In line with the best practices presented by Boyce [16], extended and descriptive responses were encouraged to explore key points raised by the interviewees in more detail. Through this interaction among participants, the goal was to achieve the co-creation of new knowledge [81]. The FGs aimed to evaluate real actions rather than intentions, emphasizing concrete anti-patterns mentioned by recruiters. This approach allowed for coverage of the aspects proposed by Morgan [56] (field, specificity, depth, and personal context) that should be observed in a FG.

In the **fifth step**, a qualitative analysis of the primary data obtained in the previous step was conducted. The transcriptions were initially read multiple times to familiarize with the data and extract excerpts from the transcription to identify emerging substantive codes in the collected data [37]. Subsequently, open and axial coding procedures from Grounded Theory (GT) were employed [76], in an inductive manner to establish a systematic analysis approach. New rounds of coding were conducted until the final form was achieved (with the final form decided in consensus between the first and last author). This dynamic and inductive approach aimed to extract explicit messages and uncover non-apparent meanings from the context [37]. The Taguette[2] facilitated the refinement and articulation of emerging codes, including their relationships as dimensions, properties, concepts, and categories [28]. However, it is essential to clarify that the intention here is not to create a theory based on the process, as is usual when using GT. Indeed, our study followed the guidelines proposed by Strauss and Corbin [76], using only some of the GT procedures to meet the research's objective, which, in this case, consists of exploring inferential and interpretative meanings to synthesize a set of anti-patterns and recommendations in the form of a mind map.

Finally, in the **sixth step**, the validity of the obtained results was rigorously ensured through participant validation, a key aspect of maintaining research authenticity [68]. Following the analysis conclusion, participants were promptly notified via email, receiving a link to the online mind map encapsulating anti-patterns and recommendations derived from the study. Additionally, participants were encouraged to share any constructive feedback they deemed valuable, fostering a collaborative and iterative refinement process. This validation step enhanced the robustness and credibility of the findings, aligning with best practices in qualitative research [21, 31].

---

[2] https://www.taguette.org

## 5 Results and Analysis

Our study examined the *attraction* and *assessment* phases within the R&S processes. Specifically, our focus centered on the step of **Collection of Applications** (refer to Section 5.1) in the *attraction* phase, and the steps of **Behavioral Assessment** (refer to Section 5.2) and **Technical Assessment** (refer to Section 5.3) in the *assessment* phase. This decision was made because these steps are the ones which require some interaction with participants. For each identified step, an exhaustive list of the most critical and recurrent antipatterns was compiled, accompanied by a set of recommendations aimed at preventing candidates from succumbing to these anti-patterns. Furthermore, Section 5.4 presents a collection of **General Recommendations** that transcend specific steps but can be applied throughout participants' professional journeys. It is noteworthy that the anti-patterns and recommendations were compiled inductively based on what has been discussed by the participants during the Focus Group (FG) sessions.

5.1 Collection of Applications

Four anti-patterns (codes) linked to the **Collection of Applications** have been identified. These <u>anti-patterns</u> (we underlined them to facilitate the visualization), along with the respective participants who articulated them, are cataloged in Table 2. Beyond elucidating these anti-patterns, actionable recommendations derived from the FGs are offered to alleviate these anti-patterns.

The first emerged anti-pattern refers to <u>Inflated Experiences and Skills</u>. Participants emphasized the problem of candidates who claim to have experience and knowledge in specific technologies but, in fact, have little to no practical experience. To provide clarity, a software engineer's career progression often includes the following levels (which can vary according to the organization, of course): trainee/intern, junior, mid-level, senior, and technical advanced roles (such as Principal, Staff Engineer, or Tech Lead). Specifically, a junior developer could have some previous experience, albeit limited, which distinguishes them from a trainee or intern. This discrepancy between what is stated in a candidate's resume (or CV) and the actual capabilities can hinder their prospects. For example, P2 mentioned "[...] the candidate states on their resume that they are a junior, but they have no experience... It's a bit of a self-sabotage because those in the field know that there's only a 1% chance that the person truly knows everything when they're still a junior with no industry experience". In agreement with P2, P1 pointed out that "[...] people sometimes lie on their resumes, but when we get to the interview and delve into the technical details, the person says, 'Oh, I only have theoretical knowledge with this tool' something we need practical knowledge of".

P14 added to this discussion by affirming that "[...] 'junior' doesn't mean someone without any experience, but rather someone with limited experience. The main difficulty we encounter is with people who are still in their under-

Table 2: Anti-patterns and Recommendations for the Collection of Applications

| Theme | Anti-patterns (codes) | Participants |
|---|---|---|
| Collection of Applications | Inflated Experiences and Skills | P1, P2, P14, P15, P16 |
|  | Lack of Alignment with the Job Profile | P1, P3, P7, P8, P13, P15 |
|  | Lack of Information in the Application | P7, P9, P11, P12 |
|  | Inappropriate Self-assessment and Positioning | P2, P3, P13 |
| **Recommendations** | | |

1) It is recommended to choose a specific technological track, such as back-end or front-end, instead of pursuing full-stack skills, aiming for deeper specialization.
2) Self-awareness and humility are essential to recognize one's limitations and adopt an appropriate posture in the professional environment.
3) Foster improvement in behavioral competencies by focusing on continuous improvement, setting clear goals, and always being committed to your objectives.
4) Demonstrate interest in both the organization and the phases and stages of the recruitment and selection process.
5) Refrain from including information on the resume that does not correspond to reality.
6) Carefully analyze the job description and tailor the application to the specific requirements of the advertised position.
7) Accurately present information about the professional contributions made in a clear and specific way, highlighting relevant results, learnings and achievements.
8) When conventional work experience is lacking, it is valid to highlight participation in bootcamps, forums, or other platforms where relevant contributions were made.
9) Develop the resume, LinkedIn profile, or GitHub in a coherent manner, outlining a trajectory that reflects a logical progression towards professional goals and desired technology.
10) Examine resumes, LinkedIn profiles, or GitHub repositories of professionals with extensive market experience to understand the paths they have taken.
11) Seek guidance from a mentor with a more established career to improve the crafting of the resume, LinkedIn profile, or GitHub repository.
12) Demonstrate proactivity in seeking resources and information to learn how to build a LinkedIn profile or resume in a cohesive and more effective manner.
13) Choose not to apply for opportunities that do not align with your purposes and values.

graduate studies or have no experience, yet they believe they fit as juniors, not understanding the difference between trainee and junior". P15 reinforced that "[...] if a person has only studied a programming language geared towards back-end development, like Java or Python, and applies for a front-end position, it doesn't make much sense". In complement, P16 explained: "You take someone who has just graduated or is early in their professional career, with a series of experiences that don't correspond to their age, and it doesn't necessarily match their actual experience. Perhaps the person thinks, 'I'll shine by including all this data', but it doesn't necessarily work out".

These previous statements highlight that inflating experiences or skills during the application is a critical anti-pattern that can negatively influence a candidate's progress in the R&S processes. As elucidated by participants, this practice creates a disconnect between stated qualifications and actual capabilities and poses a substantial risk to a candidate's progression in the R&S process. This analysis illuminates the need to refine strategies to align candidate profiles more precisely with job descriptions, optimizing the utilization of recruiters' time and ensuring a more streamlined online R&S process [61].

Another identified anti-pattern was the Lack of Alignment with the Job Profile. In this sense, interviewees mentioned the challenge of dealing with a high volume of resumes that do not match the desired profile for the position. For instance, P1 commented, "[...] we understand how eager people are to enter the job market, but many resumes are not aligned with the profile". This issue may include candidates with inappropriate experience levels, a lack of required technical knowledge, or insufficient skills. P8 reinforced that "[...] junior professionals, when applying, often do not read the job description", and P7 added, "Professionals view a job, and their profile is the opposite of that job, yet they still apply". This discussion indicates a potential challenge in resume screening, where recruiters have to spend hours sifting through profiles that are not properly aligned with the job description [55].

Still discussing the lack of alignment with the job profile, P3 also noted that "[...] people don't read the job description, especially with platforms that make it very easy to apply, like LinkedIn, where it's just one click; it's straightforward to receive a bunch of people who don't match". P7, for example, stated, "Another point I often notice is that professionals who have a long career history and have often changed fields create extensive resumes, exposing their entire career journey, including many details". Continuing in the same direction, P15 expressed the following opinion: "[...] if you work in the database field and apply for a front-end position, it doesn't make much sense." However, P13 shared a critical analysis "[...] I feel that it's a somewhat mercenary profile; they don't have a connection to the business, they don't identify with the company's values. They go where they can get the desired salary under the conditions they want".

As we can see, this discussion derived by the FGs accentuate the intricate dynamics at play in the hiring process. The challenge is not solely about aligning skills with job requirements but also about fostering a genuine connection between the candidate and the organization. The observations emphasize the need for a more strategic and thoughtful approach to job applications, encouraging a balance between technical aptitude, professional history, and cultural fit within the company [38].

Lack of Information in the Application arises from the inadequate information in junior candidates' resumes. P9 mentioned: "One thing that greatly complicates matters is the lack of information in the resumes". In agreement, P7 added, "[...] sometimes professionals simply state '.NET developer' and that's it. What did they work on? What technologies did they use?". In line with this point, P11 emphasized, "I believe the main issue, at least from what I've observed, is that people cannot create a good profile, not only on LinkedIn but also in their resumes, without providing many details about their knowledge. We are discussing people who are just starting but have experience with some technology. This absence of details about their experience and their ability to perform in the desired position complicates a thorough assessment of the candidates and often prevents them from progressing successfully to the next stages".

As we can see, discussions were held about candidates lacking self-awareness and struggling to position themselves properly based on career levels such as trainee, junior, mid-level, and senior. Participants pointed out that some candidates lack clarity about their position in their professional journey and may overestimate their skills. In this context, the last identified code was <u>Inappropriate Self-Assessment and Positioning</u>. P3, for instance, noted, "[...] the lack of self-critique is not so frequent, especially regarding certain aspects, even ignorance about one's own career or where one stands in their career, particularly nowadays when there's a trend of claiming to be a senior within a year". P13 expressed an interesting perspective, "[ ...] it's very common to receive people with a distorted financial perspective. 'I have no experience, but I want to start with the salary of a mid-level or senior professional'. This happens quite frequently, and it's quite challenging for us to mediate this".

As shown in Table 2, a series of recommendations were derived from the participants' opinions to avoid these anti-patterns in the **Collection of Applications** phase. In summary, these recommendations include selecting a specific area of specialization, self-awareness, developing behavioral competencies by focusing on continuous improvement, demonstrating genuine interest in companies and positions, and creating resumes, LinkedIn profiles, or GitHub repositories aligned with their goals. Additionally, it is advisable to seek guidance from experienced professionals, have a mentor, and refrain from applying for positions that do not align with one's objectives and values.

## 5.2 Behavioral Assessment

Regarding the **Behavioral Assessment**, five main anti-patterns were identified, as summarized in Table 3. We will now analyze these anti-patterns and, at the end, summarize potential recommendations to mitigate them.

Table 3: Anti-patterns and Recommendations for the Behavioral Assessment

| Theme | Anti-patterns (codes) | Participants |
|---|---|---|
| Behavioral Assessment | Lack of Respect for Stages, Steps, or People in the Process | P1, P2, P13, P14, P15, P17 |
| | Lack of Knowledge and Interest in the Company | P1, P4, P8, P11, P12, P14, P18 |
| | Lack of Effective Communication Skills | P7, P10, P11, P12, P14 |
| | Appropriating Team Results and Egocentrism | P3, P4 |
| | Failure to Utilize Soft Skills in the Interview | P8, P9, P11, P18 |
| **Recommendations** | | |
| 1) Be transparent, know how to recognize your limitations, and accept that it is impossible to know everything; | | |
| 2) Demonstrate a willingness to learn and grow by regularly reflecting on your experiences, embracing challenges as opportunities for growth, and showing gratitude for the guidance you receive; | | |
| 3) Research the organization, its field of operation, market tenure, and any other relevant information. Express interest in the company and demonstrate preparation for the interview; | | |
| 4) Enhance communication skills, seeking clarity and effectiveness. Utilize resources such as YouTube videos, mentoring, or even therapy; | | |
| 5) Showcase your qualities with confidence and be ready to discuss experiences, even those not directly related to the position, demonstrating versatility; | | |
| 6) Approach interviews with a calm demeanor, avoiding overwhelming pressure; | | |
| 7) Acknowledge that non-selection for a specific position does not reflect professional incompetence. Recognize that each scenario is unique; | | |
| 8) Demonstrate respect for all phases of the process and the individuals involved. Each stage has its significance, and each individual plays a significant role. | | |

The first standout anti-pattern in the context of behavioral assessment was Lack of Respect for the Stages, Steps, or People in the Process. Participants emphatically highlighted arrogance and lack of respect as unwelcome elements. They mentioned examples where candidates displayed disrespect during the selection process, mistreating interviewers or adopting a condescending attitude. For example, P1 commented, "Don't be disrespectful to the recruiter, for heaven's sake. It's very annoying and a disqualifier. I would reject the candidate without a doubt. Even if the person is technically perfect, if I sense a certain arrogance during the process, we reject them right away". On the other hand, P14 recounted a situation where a candidate said, "I won't answer your question because you won't understand it". P13 also had a similar experience: "The candidate said he wouldn't ask me a question because I'm from HR and don't know how to answer such things". The analysis conducted indicated that disregarding the stages, steps, or individuals involved in a R&S processes is a notable anti-pattern that can have adverse effects on the probability of the candidates being hired. Participants stressed the importance of professionalism and respectful behavior when interacting with interviewers and other stakeholders in the process. By avoiding behaviors such as arrogance and disrespect, candidates can increase their chances of success in the R&S processes.

Aligned with the previous finding, we also identified Lack of Knowledge and Interest in the Company as an anti-pattern. As an evidence, P1 stated: "[...] the person applies for a position in an organization without even knowing where they are applying". This lack of knowledge was perceived as a sign of disrespect toward the recruiters and the technical team. P14 emphasized that "[...] the most glaring issue, especially in terms of behavioral aspects, is that I won't approve a candidate who shows no desire to be part of the company". Even if the candidate has not conducted prior research on the company, demonstrating an interest in learning about it and being a part of it is appreciated. Nevertheless, it was observed that candidates often display disinterest, as highlighted by P11: "[...] the main issue, at least from what I've sensed, is that many candidates come with a feeling of disinterest, as if they didn't want to be here for the interview. They would say, 'I don't even know why I'm talking to you'". In a striking example, P18 explained what bothers him in interviews: "The person demonstrates that they only want the position to get away from their current job". Based on the feedback from our participants, it has been found that lack of knowledge and interest in the company is a major concern. This behavior is not only perceived as a sign of disrespect towards recruiters and technical teams but also negatively impacts the candidate's chances of getting hired. Interviewers appreciate candidates eager to learn and become a part of the company. On the other hand, candidates who display disinterest during the interview process risk being overlooked for the position.

Lack of Effective Communication Skills was another emerged anti-pattern from our analysis. In this sense, P7 explained, "[...] we understand that there are introverted people, but this is entirely different from communication skills [...]. Therefore, the development of communication skills and ongoing training in this area is crucial and will make the professional stand out wherever they are". P14 also highlighted a point that complements P7's statement: "[...] Assertive communication is different from being communicative. Assertive communication is when you have the necessary communication skills to perform your job". In this regard, P12 emphasized that in some interviews, he has to keep "nudging and prodding" participants to engage them, but even then, he receives short answers like "yes", "no", or "definitely". It is evident that lack of effective communication skills is an anti-pattern that can hinder professional growth and success [6]. Hence, the importance of effective skills was highlighted by multiple participants, who emphasized the need to communicate clearly and confidently to succeed in professional settings.

Appropriating Team Results and Egocentrism offered another valuable perspective of behavioral assessment. Participants reported instances where candidates took credit for a team's achievements, attributing collective successes to themselves. For instance, P3 highlighted, "[...] what we often see is people talking about a team's results as if they were their own, like, 'I boosted production, I did this whole front-end thing', and, man, in the end, you can tell that 15 people were involved, but they kept saying 'I' so many times, trying to convince you that they organized the whole process". In parallel to this topic of appropriating of collective results, P4 emphasized a point related to the inflated egos of these candidates who claim successful team projects solely as personal achievements: "[...] they have such a strong ego that it doesn't even allow for a structured interview". The analysis of participants' feedback revealed that there are concerns regarding the lack of recognition of teamwork, the attempt to self-promote in an egocentric manner, and a lack of humility among professionals. These findings highlight the importance of promoting and recognizing teamwork [43].

Failure to Use Soft Skills in the Interview emerged as an anti-pattern that highlights a topic that has been extensively explored in both the industry and software engineering research, which is the context of soft skills [1, 49]. Participants noted that professionals sometimes fail to leverage their soft skills and life experiences to stand out in interviews. In this regard, P9 proposed the following reflection: "[...] since you're still inexperienced, play to your soft skills, right? Use your life experiences in the meeting or interview". Building on P9's point, P11 reiterated: "[...] in terms of some soft skills, not that a beginner needs to be fully developed behaviorally, but I think it's essential to know how to discuss your trajectory, whether it's professional or related to your college experience". This analysis highlights the importance of leveraging soft skills in job interviews, especially for beginners without much professional experience [80]. Participants emphasized the need to use life experiences to stand out in interviews and discuss one's trajectory, whether professional or related to academic experience.

Regarding recommendations to avoid committing the above anti-patterns during the **Behavioral Assessment** phase, candidates should be transparent about their knowledge, demonstrate an interest in the company through prior research, develop communication skills, and be open to learning. Additionally, it is essential for candidates to know themselves, appreciate their strengths, and identify areas for improvement. This acknowledgment leaves room for candidates to recognize the importance of understanding the company's cultural perspective before applying for a position and being resilient during interviews, understanding that each process is unique and each stage serves a purpose.

5.3 Technical Assessment

Regarding the **Technical Assessment**, we identified three primary anti-patterns and four recommendations, which are summarized in Table 4 and elaborated upon below.

Table 4: Anti-patterns and Recommendations for the Technical Assessment

| Theme | Anti-patterns (codes) | Participants |
|---|---|---|
| Technical Assessment | Superficiality in Technical Terminology | P3, P16, P18 |
| | Lack of Alignment in Technical Tests | P8, P9, P11, P14 |
| | Practice of Presenting Unauthentic Work | P14, P15 |
| **Recommendations** | | |
| 1) Prioritize continuous learning and development by always looking for educational resources and learning opportunities, especially in the professional early stages; | | |
| 2) Showcase proficiency in technologies and technical approaches relevant to the position; | | |
| 3) Demonstrate the ability to explain the work carried out in the technical test, emphasizing the understanding, techniques and tools used to complete it; | | |
| 4) When doubts arise during technical tests, approach the recruitment team for clarifications. | | |

One prominent concern identified in the study pertains to the anti-pattern of Superficiality in Technical Terminology. This phenomenon manifests when candidates employ technical jargon and intricate concepts in an attempt to showcase expertise, often lacking genuine understanding or the ability to apply these terminologies effectively. As highlighted by P3, certain candidates rigidly adhere to a predefined script during technical assessments, engaging in discussions about standard topics like design patterns without a robust comprehension of their underlying purpose and utility. This issue underscores the necessity of comprehending the significance and implications of studied concepts, emphasizing the importance of a substantial foundation rather than mere verbal articulation. Moreover, as pointed out by P16, there exists a prevalent perspective that advocates for the principle of "less is more", particularly applicable to candidates in the early stages of their careers. This perspective suggests the significance of concise and focused communication, challenging the notion that verbosity equates to competence, especially in the initial phases of professional development. Addressing this concern necessitates a paradigm shift in assessment methodologies, emphasizing the ability to articulate technical terms and, more critically, a profound understanding and practical application of these concepts. As we move forward in refining hiring practices, it becomes imperative to recognize and rectify instances where candidates may lean towards a superficial display of technical knowledge.

Therefore, novice candidates are not expected to be experts in every technical aspect but rather to express a willingness to learn and actively prioritize specific areas of study. Consequently, candidates are advised to focus on transparency and avoid embellishing their skills and experience during technical interviews. They should avoid rigid scripts and aim for a flexible, context-aware approach demonstrating their field understanding.

The <u>Lack of Alignment in Technical Tests</u> also represents a anti-pattern that sheds light on the need for aligning expectations between recruiters and candidates throughout the R&S process. Examples were mentioned by the recruiters where candidates, on multiple occasions, either failed to submit the required tests without prior communication or provided responses that deviated from the specified requirements. Notably, as elucidated by P9, some candidates presumed an understanding of the test but produced responses that lacked coherence with the test's explicit specifications. This lack of alignment substantially hinders the assessment of candidates and introduces the potential for unfavorable outcomes. Additionally, P8 emphasized the need of fostering open dialogue and actively encouraged candidates to seek clarifications without hesitation, clarifying that an expectation for candidates to possess all the answers is unwarranted. This emphasis on alignment in expectations and effective communication is integral to optimizing the technical assessment process. Encouraging candidates to seek clarifications without reservation fosters an environment of collaboration and clarifies that assessments are not intended to be impossible challenges but rather opportunities for mutual understanding. The Practice of Presenting Unauthentic Work emerged as a anti-pattern shedding light on the ethical dimensions integral to professionalism. Interviewees consistently reported instances where candidates showcased projects or work they had not authentically undertaken. This behavior, as perceived by the interviewees, signifies a deficiency in ethical standards and commitment, potentially adversely impacting the candidate's overall image and assessment. P15, in particular, characterized such actions as a grave breach of professionalism, emphasizing that it exemplifies a complete lack of integrity. Addressing this anti-pattern necessitates a rigorous commitment to upholding ethical standards within the professional sphere, underlining the importance of authenticity in presenting one's work. Approaching this anti-pattern becomes urgent as organizations strive to foster environments built on trust and transparency. Recruiters should emphasize the significance of genuine representation during assessments, fostering a culture where ethical behavior is expected and integral to professional success.

To enhance the chances of success during **Technical Assessments**, candidates are advised to prioritize transparency and humility while accurately representing their skills and experiences. Demonstrating a commitment to continuous learning, especially in the early stages of one's career, appear to be crucial. Additionally, candidates should ensure a thorough understanding of relevant technologies and technical methodologies, enabling them to effectively articulate their work. When faced with uncertainties during a test, seeking guidance from the recruitment team is recommended. Embracing these recommendations contributes to a more authentic portrayal of one's capabilities and fosters an environment conducive to open communication and mutual understanding between candidates and recruiters. In conclusion, fostering a culture of transparency and commitment to genuine representation is vital in navigating the challenges highlighted in technical assessments.

In particular, early-career candidates must view continuous learning as a cornerstone to adapting to the ever-changing SE industry. A solid grasp of relevant technologies enhances a candidate's ability to effectively articulate their work and instills confidence in the evaluative process. Importantly, proactive communication with the recruitment team in moments of uncertainty ensures clarity and aligns expectations. By embodying these principles, novice candidates can contribute to a more robust and equitable assessment process, facilitating the identification and selection of individuals who possess technical proficiency and assimilate ethical and professional values essential for success in field of SE [29].

5.4 General Recommendations

Based on a thorough analysis of the outcomes of this research, we also have identified a few high-level suggestions that apply across different phases of the R&S processes. These broad recommendations can provide valuable guidance to novice candidates seeking career refinement in SE.

Optimization of LinkedIn Profiles was a critical remark by our interviewees, particularly for professional networking. To effectively showcase one's skills in LinkedIn, it is important to pay meticulous attention to certain practices. Recruiters emphasized the need for a concise "About" section, cautioning against unnecessary verbosity that might be overlooked. A brief yet informative overview of skills and notable achievements is deemed essential. Additionally, it is important to highlight strengths and provide a cohesive summary of experiences. This approach could potentially provide recruiters with a comprehensive insight into relevant skills and accomplishments, as discussed by P5. To enhance visibility among recruiters and headhunters, P1 recommended incorporating an attention-grabbing headline, strategically placed below the name, identifying key skills.

About the "Experience" section of LinkedIn, P5 and P9 suggested comprehensive detailing of professional roles, ensuring the inclusion of pertinent and cohesive information. Candidates should emphasize responsibilities, achievements, and projects that serve as evidence of their expertise. P9 mentioned that merely listing roles, such as "Worked as a developer for two months or six months" is insufficient. Instead, candidates should provide context, illustrating what they accomplished. Furthermore, the addition of keywords relevant to the desired position is essential. P1 stressed that the inclusion of keywords is instrumental in optimizing one's profile for search, enhancing the likelihood of being discovered by recruiters. Candidates should also include information about their English proficiency and other relevant skills. These additional details can be valuable for attracting companies that value specific competencies. Establishing Professional Relationships on LinkedIn was another valuable recommendation towards professional networking. To achieve this aim, candidates are advised to connect with relevant person to their field and engage in meaningful interactions. Connecting with recruiters and professionals from companies of interest and participating in groups to discuss pertinent topics is also recommended. It is important to personalize connection requests to express genuine professional interest, as advised by P8. This recommendation showcases dedication and sets candidates apart in the competitive professional landscape. Continuous interaction through commenting on relevant posts, sharing knowledge, and engaging in discussions is also important, according to P4. This approach demonstrates active interest and engagement within one's field. Strengthening professional credibility can also be achieved by recommending colleagues and eventually soliciting recommendations.

In addition, it was mentioned to maintain a <u>Professional Online Presence</u>, but it should not be the only factor considered in the assessment process. P10 and P15 recognized the value of having a digital presence, while P1 and P2 did not believe it to be a decisive criterion. P10 discussed that having a blog or an extensive portfolio of projects and a technology-focused blog can be a significant advantage, particularly for entry-level candidates. However, P2 cautioned that a strong online presence alone would not guarantee a candidate's approval. Correspondingly, P5 clarified that although a digital presence is important, the primary assessment should still be based on the candidate's technical and behavioral competencies.

Some participants recommend Customizing GitHub Profiles with care. P2 and P6 suggested keeping things simple and avoiding excessive embellishments that could detract from the main content. Additionally, P6 highlighted that while GitHub allows for specific customizations, candidates should avoid overloading their profiles with excessive graphics or animations. Instead, the focus should be on high-quality repositories, with the option to keep some projects private depending on their relevance. P11 also suggested that it's important to continually assess whether projects align with one's current level of expertise. Only projects that match a candidate's professional standing should be prominently displayed, with clarity and consistency being key in presenting one's most substantial work. Moreover, regular updates to a GitHub profile are necessary to ensure it reflects the latest skills and relevant projects.

During the discussion, the importance of Demonstrating Soft Skills was also emphasized by the participants. Collaboration, communication, genuine interest, curiosity, proactivity, resilience, leadership, emotional intelligence, and continuous knowledge updating were cited as essential skills. Effective teamwork, clear communication, and regular project updates were mentioned as essential for cultural fit and alignment with organizational values [49]. P1 discussed the significance of a collaborative attitude, as it is evaluated as one of their core values. P7 spotlighted the role of continuous learning in career development, while P4 and P8 weighted the importance of demonstrating interest and curiosity during interviews. Proactivity, resilience, and leadership were also seen as valuable skills in different situations, such as overcoming setbacks and evaluating decision-making abilities. P11 and P12 assessed candidates based on these skills, while candidates who displayed enthusiasm and a proactive approach often left a positive impression.

Still considering the domain of soft skills, emotional intelligence was widely regarded as a vital attribute for effectively managing diverse situations and interactions [57]. In this regard, P14 noted the importance of comprehending one's own values and navigating the job market with integrity intact. Candidates seeking employment should strive to exhibit emotional intelligence throughout the entire application process. Additionally, P7 stressed the importance of continually expanding one's knowledge base. A commitment to ongoing learning is highly valued, as it ensures that professionals stay abreast of industry trends and practices, positioning themselves for success [1].

## 6 Discussion

Employability is a common concern among researchers, software developers, and software organizations [4, 17, 48]. To contribute to this topic, this study has uncovered concrete evidence of anti-patterns made by novice software engineers during online R&S processes. These findings are noteworthy, as previous research has predominantly centered around the broader context of Information Technology [22, 58, 63, 77]. However, our research focuses specifically on the SE perspective and its particularities, offering a methodical approach to comprehending the obstacles that those embarking on this career path. Although our study is not confirmatory by design, it aligns with existing literature [2, 45, 46, 72], validating and extending the understanding of anti-patterns in the SE domain.

### 6.1 Implications for SE Research

To guide this work, we defined the following research question: "*What anti-patterns are committed by early career professionals undergoing the online R&S processes for positions in the field of Software Engineering?* ". After investigating this inriquiry, our hypothesis concerning the existence of antipatterns was confirmed, as we have identified several recurring negative actions that hinder the success of novice professionals in SE industry. Through empirical research, we have also identified 12 anti-patterns that can impact candidates during the R&S process. By understanding and addressing these issues, we also offered 31 actionable recommendations that may help streamline the hiring process, aiding both job seekers and companies engaged in effective R&S procedures.

Concerning the **Collection of Applications**, Sonmez [72] also mentioned the challenges that applicants might encounter at the application stage. His conclusions favored the importance of steering clear of overly lengthy and disorganized resumes, neglecting personal branding, and failing to distinguish oneself from the competition. In alignment with this perspective, our identified anti-pattern, Lack of Information in Application, resonates with these observations, spotlighting the issue of insufficient or excessively detailed information in the application process. The emphasis should be on conveying pertinent details that succinctly showcase the candidate's skills, experiences, and achievements. Moreover, the importance of personal branding cannot be overstated, as it serves as a distinctive element that helps candidates establish a memorable and differentiated identity in the eyes of recruiters [20].

Regarding the **Behavioral Assessment**, the results indicated that candidates commonly exhibit critical anti-patterns such as disregarding the process's stages, steps, or people, lacking knowledge and interest in the company, displaying ineffective communication skills, exhibiting egocentrism and team result appropriation, and failing to utilize soft skills during interviews [2, 46]. Noteworthy is the anti-pattern Lack of Respect for the Process's Stages, Steps, or People in Selection, which aligns with challenges identified by Ahmed and Capretz [1], emphasizing the need for candidates to address soft skills for sustained professional growth in the ever-evolving SE landscape.

When it comes to the **Technical Assessment**, candidates have access to a wealth of resources, including books, online guides, practice websites, and experience reports. Indeed, these tools also offer valuable knowledge and strategies to help navigate coding challenges [45]. While these tools also offer valuable knowledge and strategies to help candidates navigate coding challenges, our study complement them by pinpointing recurring pitfalls and offering tailored guidance. Hence, novice engineers can fortify their preparation with wisdom that directly addresses potential areas of weakness, such as Superficiality in Technical Terminology. This nuanced approach contributes to a more comprehensive and targeted preparation strategy, enhancing early-career candidates' ability to navigate technical assessments successfully.

Our investigation additionally yielded **General Recommendations** that exhibit applicability throughout various stages of the R&S processes. These overarching directives offer pragmatic guidance for novice engineers aspiring to refine their professional trajectories in SE. These recommendations revolve around the enhancement of LinkedIn and GitHub profiles and the proficient demonstration of soft skills. It is noteworthy to observe that the relevance of social networks in the context of SE has already been highlighted by SE researchers [47, 75], but it is still considerable scarce. Furthermore, the importance of soft skills has been a subject of considerable importance over the last years [48], including the industry demand [49].

6.2 Key Takeaways for Industry

In summary, this study identifies common anti-patterns in the online R&S process and provides recommendations from recruiters to help candidates avoid repeating these errors. The findings offer organizations a strategic lens to optimize talent acquisition by addressing specific anti-patterns identified in the SE recruitment process. Implementing these recommendations can refine recruitment procedures, mitigate risks associated with suboptimal hiring decisions, and enhance overall efficiency in candidate selection. As discussed by More, Stray, and Smite [52], competing for talents requires a conscious effort to increase employee empowerment and engagement. On the other hand, early-career SE candidates can boost their success rates by heeding the study's recommendations. Understanding and avoiding the identified anti-patterns enables candidates to present themselves more effectively during the R&S processes, increasing their chances of employability.

Therefore, this paper does not exclusively target academics since it enables transfer to industry by providing anti-patterns and recommendations to mitigate them. This holistic perspective embraces the interconnected nature of behavioral and technical competencies, guiding industry stakeholders toward successful talent management strategies. To ease the comprehension, the results have been synthesized in a concise and interactive mind map, available at the following link: https://gesid.github.io/papers/swe-novice-rs. This visual representation simplifies the understanding of the identified anti-patterns and recommendations for different steps of the R&S process. In addition, this output may serve as an executive summary tailored for industrial readers.

Lastly, as discussed in the sixth step of our methodological procedures (see Section 4), our invited participants were also asked to validate this mind map via email. Overall, the feedback was overwhelmingly positive. For instance, P5 remarked, "I loved the mind map, it's very comprehensive and summarizes everything we discussed". P1 expressed their gratitude, saying, "I'm glad to have contributed to the research". Lastly, P9 concluded, "[...] I think the material turned out great; it's a good guide and seems very useful!". These statements indicate that the participants received the results favorably. However, we know the importance of weighing how much their potential introspection has influenced their responses.

## 7 Limitations

Although this study has successfully identified a series of anti-patterns and proposed helpful recommendations for early-career professionals navigating the online R&S process for SE roles, it is important to acknowledge the limitations that may have impacted it. One potential limitation is the relatively small sample size of participants in the Focus Groups (FGs). While the results we obtained are significant, it is worth noting that a more comprehensive perspective could have been attained with a larger sample size. Hence, our findings call for more research on the intersection between HR Management and SE, including the development of guidelines, processes, and artifacts.

In terms of data collection, the participants were interviewed in their natural work settings (in a non-controlled environment), which could potentially influenced their engagement. However, since the subject matter was already a component of their daily responsibilities, this potential issue was mitigated. To reduce any possible impact, we scheduled interviews well ahead of time and provided a detailed explanation of the study's objectives. Despite these measures, there is a possibility that participants may have altered their behavior and responses as a result of being under observation from their peers. In this sense, it is worth noting that the speeches cited in this body of work have been translated from Brazilian Portuguese to English, and despite the considerable effort devoted to ensuring accuracy, there is a possibility that certain nuances may have been lost in translation due to differences in vocabulary. All the data underpinning this study (informed consent form, participant characterization questionnaire, interview script, mind map, and full FGs transcriptions) is openly available via our supporting repository [67], ensuring transparency and accessibility.

Furthermore, it is important to mention that this study is not intended to be all-encompassing or provide exhaustive explanations for all existing anti-patterns. As each company has its own distinct characteristics, the R&S processes may differ. These anti-patterns may vary depending on the organization, taking into account factors such as business segment, maturity level, and business model. To overcome this obstacle, this research relied on a methodological approach that involved inviting a diverse group of experienced recruiters, allowing for the identification of anti-patterns from various perspectives.

To ensure a representative and qualified composition of our FGs, we established two participation criteria: i) currently working in a technology-based company and ii) a minimum of one year of experience in R&S of software engineers, whether as a recruiter, tech recruiter, manager, tech lead, or other relevant role. On average, participants have 5.7 and 3.3 years of experience in "traditional" and online R&S, respectively. The most experienced participant has 20 years in "traditional" and 11 years in the online context. While two participants have no experience in "traditional" R&S, they have two years of experience in the online R&S. Our participants came from a range of academic backgrounds, including technology, psychology, and engineering. Notwithstanding the emphasis placed by the FG facilitator on early-career candidates, participants may have occasionally incorporated a systemic view due to their previous experience with other profiles, including more experienced SE professionals.

## 8 Concluding Remarks

In today's professional landscape, online Recruitment and Selection (R&S) processes are commonplace, particularly within the Software Engineering (SE) industry. Nevertheless, many early-career professionals may make common mistakes due to their lack of experience and unfamiliarity with anti-patterns, potentially minimizing their chances of securing a position.

In this context, this study aimed to investigate the online professional R&S journey of early-career software engineers from the perspective of recruiters. The goal was to identify anti-patterns and provide potential recommendations. To achieve the desired outcome, we conducted an empirical and qualitative research using six Focus Groups (FGs) with experienced recruiters. The methodological procedures involved six main steps. Firstly, a minor literature review was conducted to understand the fundamental concepts and definitions of theoretical elements connecting SE and Human Resources (HR) in the context of R&S. Next, an interview script was defined for collecting primary data through FGs. Participants were then invited to join the FGs in the third stage. The fourth stage focused on conducting the FGs with confirmed participants. In the fifth stage, a qualitative analysis of the primary data obtained earlier was performed using open and axial coding procedures. Finally, the results were validated with the participants in the sixth stage.

In terms of results, we were able to identify 12 recurring anti-patterns that hinder novice candidates' progress in the online professional hiring process. In addition, we came up with 31 recommendations to help candidates mitigate these anti-patterns. We also developed a list of General Recommendations that covers high-level aspects of building a professional career, rather than just one specific phase. These results were synthesized in the form of a comprehensive mind map (available at https://gesid.github.io/papers/ swe-novice-rs).

As we can see, this research makes a contribution to both industry and academia. Firstly, it offers a resource for the SE industry by mapping anti-patterns and providing actionable recommendations. Early career candidates, in particular, can benefit greatly from the nuanced understanding gained of anti-patterns to avoid during the online R&S processes. From the company's perspective, our analysis contributes to optimizing their talent acquisition strategies. Secondly, our study advances the broader academic discourse by delving into the intersection of HR and SE, an underexplored research domain in the context of cooperative and human aspects of SE. By bridging together these disciplines, we contribute to developing a more holistic understanding of the complexities and intricacies involved in the R&S of software engineers. This interdisciplinary approach offers theoretical and practical implications that can inform future research endeavors in HR and SE domains. Additionally, our study recognizes the behavioral and technical dimensions inherent in the R&S processes, aligning with the growing emphasis on considering social aspects in SE research. This recognition enhances our understanding of the sociotechnical dynamics in the R&S and their impact on forming highly skilled software development teams.

Finally, in terms of future work, we aim to enhance our research by delving into the candidates' perspective, and gaining discernment into their experiences in R&S. Additionally, we may explore the perception of experienced developers, leading to a discussion on the contrasts and changes that emerge with participant maturity, including their view on the possible anti-patterns practiced by R&S professionals. Moreover, additional investigations could be conducted to expand the interactive online material. For instance, by linking individual recommendations to complementary sources (*e.g.*, blog posts and YouTube videos) that provide more practical details, or by supplying concrete example scenarios. Future work could also involve collaborations and replications to further explore an international (or cross-country) perspective aiming to improve the generalizability of our findings.

**Data Availability Statements**

This study relies primarily on qualitative data from focus groups. All the supplementary material is available at https://doi.org/10.5281/zenodo. 10436033. It includes: a) Informed consent form; b) Participant characterization questionnaire; c) Interview script for the focus groups; d) Mind map (in .pdf format); e) Focus groups transcriptions.

**Acknowledgments**

We would like to express our gratitude to the participants for their availability and to the reviewers for their valuable suggestions. This research was partially funded by CAPES - Financing Code 001, as well as CNPq processes 314797/2023-8 and 312275/2023-4. Additionally, we would like to acknowledge the support of the following institutions: Universidade Federal do Ceará (UFC), Universidade Federal do Amazonas (UFAM), Pontifícia Universidade Católica do Rio de Janeiro (PUC-Rio), and Universidade Federal do Cariri (UFCA).

**Statements and Declarations**

- The authors have no relevant financial or non-financial interests to disclose;
- The authors have no competing interests to declare that are relevant to the content of this article;
- All authors certify that they have no affiliations with or involvement in any organization or entity with any financial interest or non-financial interest in the subject matter or materials discussed in this manuscript;
- The authors have no financial or proprietary interests in any material discussed in this article.